\documentclass[journal]{IEEEtran}
\usepackage{cite}
\ifCLASSINFOpdf
\else
\fi

\usepackage[cmex10]{amsmath}
\usepackage{array}
\usepackage{graphicx,cite,bm,algorithm,algorithmic,amsfonts,amsthm,amsmath,pifont,url}
\usepackage{color}
\usepackage{epstopdf}

  \theoremstyle{plain}
  \newtheorem*{points*}{Main points}

\begin{document}
\title{Power Systems Without Fuel}
\author{Joshua A.~Taylor,~\IEEEmembership{Member,~IEEE}, Sairaj V.~Dhople,~\IEEEmembership{Member,~IEEE}, and Duncan S.~Callaway,~\IEEEmembership{Member,~IEEE}
\thanks{J. A. Taylor is with the Department of Electrical and Computer Engineering at the University of Toronto, Toronto, Canada.  E-mail: \texttt{josh.taylor@utoronto.ca}.}
\thanks{S. V. Dhople is with the Department of Electrical and Computer Engineering at the University of Minnesota, Minneapolis, MN, USA.  E-mail: \texttt{sdhople@umn.edu}.}  
\thanks{D. S. Callaway is with the Energy and Resources Group at the University of California at Berkeley, CA, USA. E-mail: \texttt{dcal@berkeley.edu}.}
}

\maketitle
\begin{abstract}

The finiteness of fossil fuels implies that future electric power systems may predominantly source energy from fuel-free renewable resources like wind and solar. Evidently, these \emph{power systems without fuel} will be environmentally benign, sustainable, and subject to milder failure scenarios. Many of these advantages were projected decades ago with the definition of the \textit{soft energy path}, which describes a future where all energy is provided by numerous small, simple, and diverse renewable sources. Here we provide a thorough investigation of power systems without fuel from technical and economic standpoints. The paper is organized by timescale and covers issues like the irrelevance of unit commitment in networks without large, fuel-based generators, the dubiousness of nodal pricing without fuel costs, and the need for new system-level models and control methods for semiconductor-based energy-conversion interfaces.

\end{abstract}
\begin{IEEEkeywords}
Optimization, Power electronics, Power system operation, Renewable energy, Soft energy path.
\end{IEEEkeywords}

\section{Introduction}
\IEEEPARstart{E}{ventually}, whether from a deliberate shift to renewables or the depletion of planetary fossil fuel sources, major portions of power systems will run without fossil fuels. There is now extensive literature on this subject, see, e.g.,~\cite{mackay2008sustainable,tester2012sustainable}, most of which focuses on the advantages and disadvantages of renewable energy sources. Here we also consider this scenario, but focus on the combined consequences of using renewables and eliminating fuel-based generators; we fully define our scope in Section~\ref{setup}. We believe that it is important to discuss these issues both for the limiting case of exclusively renewable power production and also under mostly renewable power production, in which case planning and operational practices should reflect the large majority of renewables instead of a small minority of fuel-based generators.


Much of the salient physics of present-day power systems are attributable to fuel-based generators, which are most cost-effective at large unit sizes; for example, thermal limits constrain power production schedules, and synchronous machine rotor inertias dominate transient stability. While the addition of intermittent and distributed renewables will change the form of power systems, so will removing the fuel-based generators that account for much of its current character. Many of the consequences of removing fuel-based generators are well studied, for example the loss of system inertia and the need to replace generator reserves with storage and demand response. However, it is our perception that many important issues have not been thoroughly discussed, such as the need for new optimal power flow objectives to replace fuel costs and the attendant need for alternative economic mechanisms to nodal pricing. In some cases, removing fuel-based generators can result in significant benefits like the elimination of the unit commitment problem and the viability of transitioning to a predominantly DC infrastructure. We address such issues in a unified fashion by surveying existing discussions, identifying challenges, and suggesting new directions where appropriate. 

\begin{figure*}[t!]
\centering \includegraphics[scale=0.9]{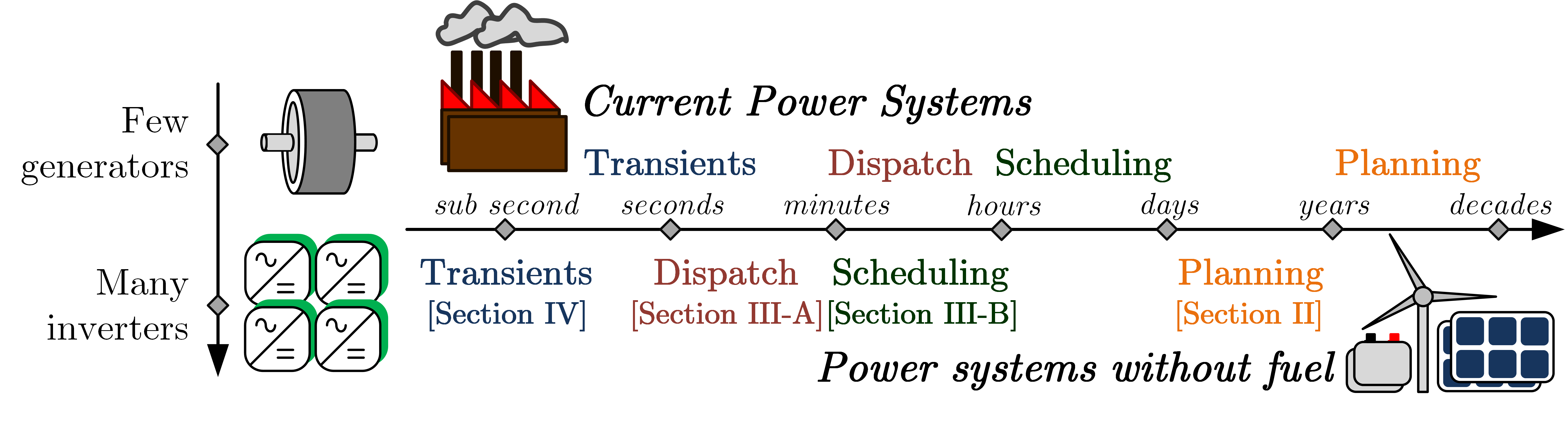}
\caption{No-fuel power systems will have fundamentally different spatio-temporal characteristics. Time scales governing operations and control will shrink, and energy-conversion interfaces will be distributed in form and function.}
\label{fig:TimeScales}
\end{figure*}

Two key motifs of this paper are that (see Fig.~\ref{fig:TimeScales})
\begin{itemize}
\item all timescales are shrinking, and
\item a few large components will be replaced by numerous small components.
\end{itemize}
These changes have a number of salient consequences. For instance, renewables can be installed and maintained more quickly than fuel-based generators, bulk generation need not be scheduled days in advance to accommodate stringent startup and shutdown constraints, and the loss of mechanical inertia implies that transients will be dominated by the dynamic behaviors of many small, fast power-electronic energy-conversion interfaces. We partially structure our discussions around timescales by first considering long-term planning, then steady-state operation, and then transient dynamics. While so doing, we additionally highlight the emergence and disappearance of new and old couplings. For example, renewable intermittency couples hourly steady-state dispatch decisions to regulation requirements. Similarly, wind and solar power sources consume negligible quantities of water compared to the cooling needs of fuel-based generators, drastically reducing the power system's dependency on the water infrastructure. 


\subsection{The soft energy path and the status quo}\label{soft}
The rationale for a fully renewable energy infrastructure has long been well established; we now briefly summarize its beginnings and provide a few modern examples. In 1976, Lovins described the \textit{soft energy path} as a future in which societal energy needs are met by numerous small, simple, diverse technologies that rely on renewable energy sources~\cite{lovins1976energy,lovins2013reinventing}. The soft energy path stands in contrast to the \textit{hard energy path}, wherein energy is provided by a few large, resource intensive, complex technologies which inevitably induce caustic political dependencies and are prone to expensive physical failures. The case for the \textit{soft energy path}---which we note, does not include nuclear power---has been restated and reaffirmed in many ways over the past four decades; for instance, the capability of wind and solar to fulfill all of our energy needs is addressed in~\cite{Jacobson2011all1}. Distributed generation~\cite{Ackermann2001def} and microgrids~\cite{lasseter2002micro} are two related architectural paradigms where small, renewable technologies meet energy needs locally. Another compelling argument for the soft energy path is that \emph{unused} wind and solar energy are lost while fossil fuels may be indefinitely ``... [left] ... in the ground for emergency use only''~\cite{lovins1976energy}.

It may well be some time before continental-scale portions of power systems are fully on the soft energy path. However, there are a number of encouraging small-scale examples, a few of which we list below:
\begin{itemize}
\item Scotland exceeded its target of 31\% renewable electricity production by 2011, and has established targets of 50\% and 100\% by 2015 and 2020, respectively~\cite{scotland2020}.
\item Following a destructive fire, the Caribbean island of Bonaire now aims to generate all of its electricity from renewables~\cite{bonaire100}, about 45\% of which will be wind and the remainder biofuel or biomass. Geographical remoteness makes local renewable production highly pragmatic for small island power systems~\cite{ilic2013engineering}.
\item Following a nuclear disaster in 2011, the Fukushima prefecture in Japan established 100\% renewable electricity production as their target for 2040~\cite{WorldFutureCouncil}.
\item Portugal, which has no established companies in fossil or nuclear fuel production~\cite{Reiche2004eu}, aims to have 60\% renewable electricity production by 2020~\cite{100Portugal}.
\end{itemize}

\subsection{Setup and organization}\label{setup}
In this paper, we attempt to broadly discuss the implications of fully transitioning electric power systems to a no-fuel future. We implicitly take the virtues of the soft energy path as given and assume almost $100\%$ renewable power production to be a viable future for a large share of power systems. In this regard, our main objective is to better understand power systems without fuel as a limiting case to help plan for an energy future that is potentially very different from the present.

By renewable, we primarily connote wind, solar, wave, and hydroelectric power production, but also allow for sustainable low-fuel, low-pollution resources like geothermal energy and fuel cells. Because we are interested in both all-renewable and almost all-renewable power production, we allow for occasional instances of these resources that contradict our setup. For example, pumped hydro and concentrating solar power plants often use synchronous generators and can be in the hundreds of MW, which is much larger than most renewables; however, both technologies still have very low marginal costs and are amenable to direct current configurations. Note that the scope of the paper does not encompass nuclear generation or biomass and biofuels, the latter of which may remain useful for some non-grid connected types of energy consumers like long-distance air travel~\cite{mackay2008sustainable,Searchinger2008US,Nigam2011liquid}.

Throughout our discussion, we implicitly assume the existence of large stores of flexibility, by which we mean the ability to match supply and demand on all timescales and maintain stable voltages throughout the power system. The central challenge in replacing conventional fuel-based generators with renewables is finding inexpensive sources of flexibility. Presently, flexibility is achieved via reserves and regulation. Reserves balance supply and demand in the face of unpredicted load variations and contingencies, i.e., component failures. Regulation, which includes automatic generation control, automatic voltage regulation, and droop control, balances supply and demand on faster timescales and maintains stable voltage frequencies and magnitudes throughout the power system. The reader is referred to~\cite{kundur1994stab,WoodWoll2} for comprehensive coverage of these topics. Presently, most of these services are provided by fuel-based generators. Replacing these generators with renewables simultaneously increases the need for flexibility and removes the main source of flexibility. A core obstacle is that most new sources of flexibility are still quite expensive.

Because we are taking a long-term view in this paper, we neglect the challenge of financing the procurement of new sources of flexibility, and rather assume that sufficient flexibility can be provided by the following non-fuel-based mechanisms.
\begin{itemize}
\item Energy storage devices like batteries, flywheels, and pumped hydro~\cite{Castillo2014survey}. A generic model of storage is given by equations~\eqref{eq:storage0}--\eqref{eq:storage2} in Section~\ref{LS}.
\item Flexible loads enrolled in demand response programs~\cite{callaway11EL,dietrich2011side}. Several recent studies have shown that flexible load aggregations can provide similar services to (and are well-modeled by) storage~\cite{mathieu2013loads,taylor2013FlexCDC,hao2014thermo}. For this reason, we will generically use the term \textit{storage} to refer to devices like batteries and flywheels \emph{and} to virtual storage from aggregations of flexible loads.
\item \textit{Spilling} renewable power by operating below capacity limits. The maximum power output of a wind or solar power producer depends on prevailing ambient conditions which are inherently random. Renewables are largely dispatchable up to their maximum power outputs~\cite{bianchi2006wind,patel2006windsolar}.
\item Power electronic devices. Converter interfaces for renewables, storage~\cite{carrasco2006pes}, and direct current systems~\cite{arrillaga1988HVDC,Demetriades2009hvdc} and FACTS devices~\cite{hingorani1999facts,acha2004FACTS} can provide voltage support and regulation on fast timescales. 
\end{itemize}
The remainder of the paper is organized as follows. In Section~\ref{plan}, we discuss planning and some broader issues like couplings to other infrastructures and DC systems. We move to the steady-state timescale in Section~\ref{steady}, where we discuss the scheduling problems of unit commitment and load-shifting, real-time dispatch, and the economics of power systems without fuel. We discuss dynamics and transient stability in Section~\ref{stability}. Figure~\ref{fig:TimeScales} illustrates the paper's organization within the motif of evolving spatiotemporal characteristics.

\section{Long-term planning and broader issues}\label{plan}
In this section, we discuss planning with renewables and the broader issues of direct versus alternating current and couplings with other infrastructures.

\subsection{Economies of scale and geographic constraints}\label{planscale}
This section focuses on basic differences between planning with fuel-based generators and with renewables and storage. We structure our discussion around the following generic optimal planning problem.
\begin{equation}
\begin{array}{rl}
\underset{x,y}{\mathrm{minimize}} &\underset{\sigma}{\mathbb{E}}\left[\displaystyle \sum_{i\in\mathcal{C}} c_i(x_i,\sigma) +f(x,y,\sigma)\right]  \\
\mathrm{such\,\,that} & 0\leq x_i\leq \overline{x}_i,\quad x_i\in\mathbb{Z},\\
& y\in \mathcal{P}(x).
\end{array}
\label{eq:plan}
\end{equation}
Here, $x_i$ represents the decision to install component $i\in \mathcal{C}$, where $\mathcal{C}$ is the set of candidate components, and $y$ is the vector of operational variables, for example, real and reactive power flows and voltage. The first constraint on $x_i$ represents the quantity installed, and the second constraint introduces a discrete requirement, e.g., that the number of wind turbines must be integer valued. The feasible set of operational variables depends on the installed components via $\mathcal{P}(x)$. The first term in the objective is the sum of installation costs, and the latter, the total operating cost. Because installation can span weeks to decades, the expectation of the cost is taken over $\sigma$, a vector of random variables representing uncertain future information like load growth and material and fuel costs.

Economies of scale reduce fuel-based generator costs as installed capacity increases. On the contrary, wind turbines and PV inverters are typically available at smaller ratings, often one to two orders of magnitude less than fuel-based generators in terms of watts per facility. For example, in 2013, the average U.S. coal, gas, and nuclear plant capacities were approximately $272\,\mathrm{MW}$, $85\,\mathrm{MW}$, and $104\,\mathrm{MW}$, respectively~\cite{EIA2012capacity}. On the other hand, the Vestas V164 has the largest capacity of any individual wind turbine at $8\,\mathrm{MW}$~\cite{vestas164}, and the average installed residential-scale PV system was $6.2\,\mathrm{kW}$ in 2012~\cite{Sherwood12}. While constructing a fuel-based generator takes around five years, wind turbines typically take a few months, and photovoltaic systems and many storage technologies need only be interconnected. Fuel-based generators are built to operate for several decades. Wind turbines are typically designed for a twenty-year lifetime~\cite{Martinez2009life}, while PV panel components can be expected to last anywhere from five years to two decades~\cite{amarantunga2008lifetime,Voroshazi2011life}. PV can be incrementally installed on daily timescales, which should be reflected in scheduling and dispatch routines.


Renewables are subject to different geographical constraints than fuel-based generators. Typical generation planning constraints may include minimal distance from population centers and proximity to water sources for cooling. On the other hand, renewables are geographically constrained by weather characteristics such as in the cases of offshore wind and desert solar power plants, which may entail significant new transmission investment~\cite{Willis2000DG,piwko2005wind}. Similarly, accommodating rooftop PV systems in existing distribution networks calls for an upgrade in distribution-system infrastructure. In the short-term, this implies that renewables may require more transmission expansion than fuel-based generators. In the long-term, this may incentivize large power consumers to locate in energy dense regions. There are few (if any) geographical constraints on non-hydroelectric storage.

In economic terms, the above-described smaller unit sizes make planning renewable energy installations considerably less \textit{lumpy} than fuel-based generators; rather than infrequently building large generators over long intervals, wind turbines and solar panels can be installed incrementally over shorter time periods. We summarize the resulting differences in the context of the optimization problem in~\eqref{eq:plan} below.
\begin{itemize}
\item Planning renewables has lower computational complexity than fuel-based generators. For instance, because wind turbines and solar panels come in finer size increments than fuel-based generators, the sets of possible farm sizes are better approximated by continuous intervals. In~\eqref{eq:plan}, this means increasing the number of sizes a candidate component can take on, $\overline{x}_i$, and decreasing the sensitivity of the cost, $\sum c_i(x_i,\sigma)+f(x,y,\sigma)$, and feasible operating range, $\mathcal{P}(x)$, on the installation variable $x_i$. This improves the quality of continuous approximations, and in turn, the performance of integer programming techniques like branch-and-bound~\cite{wolseyIP, Schrijver1998LPIP}, making optimal plans easier to obtain. 
\item Renewables need shorter planning horizons than fuel-based generators. This leaves less time for predictions at the start of the project to deviate from reality once completed, for example due to unexpected load growth. In (\ref{eq:plan}), this means reducing the variance of the random variable $\sigma$. While this may not affect the expectation of the objective, it decreases the risk by reducing the likelihood of the expected cost diverging from the realized cost. Consequently, there is less uncertainty from long planning horizons with renewables, decreasing the risk of wasted investments. 
\item In power systems that conduct generation expansion through market mechanisms rather than the centralized solution of mixed-integer programs (see, e.g., \cite{smeers2005imp,fu2007market}), having more numerous, smaller market participants more nearly satisfies the underlying theoretical assumptions of competitive markets \cite{mas1995microeconomic}. In particular, renewable producers will be closer to \textit{price-takers}, thus enhancing competition and easing market entry for new firms.
\item On the other hand, power systems that rely on market signals to incentivize capacity expansion will likely rely increasingly on \textit{capacity payments}, i.e., payments for generators to be available to deliver power in the future, and less on coordinated \textit{spot markets} for energy, which currently compensate producers at the marginal cost of the highest accepted bid. ``Energy only" wholesale markets (i.e., markets with only spot energy prices and without capacity payments) such as those in Texas and Alberta will probably need to be re-designed to ensure that capacity expansion continues to meet reliability requirements in the face of potentially reduced energy market revenues.  
\end{itemize}


\subsection{Coupling to other infrastructures}

In this section, we address dependencies on other infrastructures. Removing fuel-based generators decouples power systems from other infrastructures because in the envisioned no-fuel future:
\begin{itemize}
\item fuel does not need to be mined and transported via ship, rail, road, or pipe;
\item significantly less waste (e.g., ash from coal power plants) needs to be disposed of; and
\item water is not necessary for cooling, as it currently is for thermal power plants.\footnote{We remark, however, that in some cases solar thermal generation requires significant amounts of water~\cite{macknick2011review}.}
\end{itemize}
These decouplings further reduce planning complexity by removing considerations such as collocating fuel-based generators and desalinization plants.

Power systems without fuel will however be more coupled to the transportation and communication infrastructures. Coupling power systems to transportation through electric vehicles and trains will increase complexity, but at the benefit of decoupling transportation from the fuel infrastructure. Leveraging smart meters and employing more real-time control~\cite{gungor2011smart} enhances the achievable performance of power systems while introducing security and privacy vulnerabilities~\cite{mcdaniel2009smart}.


\subsection{AC versus DC and the relevance of frequency}\label{acdc}
Since the so-called ``War of the Currents'' of the 1880s, AC has been a core feature of power systems around the world;~\cite{munson2005edison} provides an excellent historical account. Since then, the question of which would be better today has often been posed, cf.~\cite{hammerstrom2007ACDC}. We now do so again for power systems without fuel. For cohesion, we contain our entire discussion of this issue to this section despite certain aspects like stability informing faster timescales discussed elsewhere in the paper.

First, recall the two main reasons for using AC instead of DC.
\begin{itemize}
\item Transformers only work with AC and until recently were the only practical solution for long-distance transmission.
\item Synchronous machines can directly synchronize with AC power systems.
\end{itemize}
Today, power electronics offer alternative mechanisms for varying current to voltage ratios that are compatible with both AC and DC~\cite{erickson2001fundamentals}, negating the first point above. Absent synchronous generators, synchronous and induction motors are the only remaining major classes of devices that are sinusoidally excited. Electric motors make up approximately 45\% of global electric power consumption, 23\% of which are synchronous for a total of approximately 10\% of the global load, a clear minority~\cite{waide2011energy}. Almost all other loads merely tolerate AC because the unwanted effects of 60Hz oscillations are imperceptible (e.g., lighting) or because they are equipped with adaptors that locally covert to DC (e.g., computing). 

The major shortcoming of DC relative to AC is at the protection level; AC faults are easily cleared because the current crosses zero 120 times per second, but DC current is constantly nonzero, leading to potentially dangerous arcing when circuits are opened. Viable protection schemes exist, especially for low-voltage settings~\cite{Sannino2007DC}, but they are more expensive than their AC counterparts. In high-voltage settings, new technologies like multi-level voltage source converters offer promising solutions \cite{iravani2012vsc,merlin2014fault}. Since conceptual solutions already exist, we regard DC protection as a surmountable obstacle that will become more economical with future research.


The features that made AC indispensable for the power systems of the last century, voltage conversion via transformers and machine synchronization, respectively can be achieved by other means (power electronics) or serve little purpose in power systems without fuel. We list several distinct differences between AC and DC transmission systems below.


First, high-voltage DC transmission is an established technology~\cite{arrillaga1988HVDC,johnson2007abc,Demetriades2009hvdc} that is superior to AC in terms of
\begin{itemize}
\item only needing one rather than three phases,
\item more efficient wire utilization because the RMS and peak currents are approximately the same,
\item providing enhanced stability via additional controllability and physical decoupling~\cite{andersson1993hvdc,cao2012wide,azad2013mpc},
\item and requiring reduced communication due to amenability to decentralization~\cite{taylor2014DC,azad2015LCC}.
\end{itemize}

DC is also viewed as a conceptually superior option for small power systems such as industrial sites~\cite{baran2003dc,Sannino2007DC} and microgrids~\cite{miura2010dc,castilla2011hier} because of
\begin{itemize}
\item improved efficiency due to decreased losses between DC sources and loads,
\item absence of synchronization and reactive power considerations,
\item and improved isolation from faults and failures on the main grid.
\end{itemize}

Finally, operating DC systems in steady-state is mathematically simpler than AC systems. This was part of the reason behind Edison's opposition to AC. The modern pragmatic implication is that DC systems lend themselves to easier computations, further reducing the operational complexity of power systems. In particular, convex relaxations of DC power flow~\cite{gan2014dc} are exact under more general conditions than convex relaxations of AC power flow~\cite{low2014partI,low2014partII}.\footnote{To avoid confusion amidst the growing presence of DC technologies, we suggest that the phrase ``DC power flow'' no longer be used to refer to linearizations of AC power flow.} This means that solving the standard nonconvex optimal power flow is tractable in the DC case for a wider range of systems than the AC case. 

\begin{points*}
Planning renewables introduces new transmission requirements. However, planning renewables is an easier problem than planning fuel-based generators because
\begin{itemize}
\item renewables and storage devices come in smaller units, improving the accuracy of continuous approximations to planning problems and hence decreasing the difficulty of the associated integer programs; and
\item they can be installed over shorter time periods, decreasing uncertainty from long planning horizons.
\end{itemize}
The following two considerations suggest that power systems without fuel can attain superior performance to that of current power systems.
\begin{itemize}
\item Power electronics and the absence of synchronous generators make DC a viable paradigm. 
\item With the exception of communications, power systems without fuel will be less coupled to other infrastructures, reducing complexity and enhancing robustness.
\end{itemize}
\end{points*}

\section{Steady-state Operation}\label{steady}
In this section, we discuss the effects of removing fuel-based generators on steady-state operations where bulk supply is matched to demand.

\subsection{Scheduling}\label{schedule}


The two main scheduling problems in power system operations are
\begin{itemize}
\item unit commitment, the problem of selecting when fuel-based generators are online and offline, and
\item load shifting, the problem of moving energy in time via storage.
\end{itemize}
Eliminating fuel-based generators and adding renewables and storage eliminates unit commitment and increases the system's load-shifting capability. Impacts of renewable intermittency are more pronounced on real-time dispatch, which we address in Section~\ref{dispatch}.

\subsubsection{Unit commitment}
Unit commitment has long been a central part of power system operations~\cite{sheble1994unit,padhy2004UC,WoodWoll2}. Unit commitment is challenging because a fuel-based generator can take hours to a day to switch between on and off states. Consequently, the attendant scheduling problem requires binary variables to indicate when fuel-based generators are on and off, and unit commitment solutions must be obtained hours to days in advance of real time. The resulting combinatorial problems are often posed as mixed-integer programs~\cite{carrion2006unit}, which, despite the development of powerful heuristics, are NP-hard and hence difficult to solve~\cite{wolseyIP,Schrijver1998LPIP}. Generator reserves---idling generators kept online as insurance against failures---must also be scheduled in advance, and are usually incorporated into unit commitment routines~\cite{galiana2005pst,WoodWoll2}.

Unlike fuel-based generators, wind turbines and hydroelectric generators can go from off to maximum output in minutes due to the absence of thermal constraints~\cite{ackerman2005wind}. Solar panels and distributed energy-storage devices that are interconnected through power electronics can engage and disengage almost instantaneously~\cite{patel2006windsolar}. Without fuel-based generators, the most stringent of these startup and shutdown constraints amounts to a few minutes, which is essentially real-time with respect to current power system operational practices. Therefore, a major perk of eliminating fuel-based generators is the elimination of the unit commitment problem and the concomitant need to procure reserves long in advance, which in turn has the following benefits:
\begin{itemize}
\item The computational complexity of power system operations decreases because combinatorial scheduling problems, i.e., integer programs, do not need to be solved.
\item The uncertainty resulting from scheduling fuel-based generators hours to days in advance disappears. Instead, renewables and storage can be brought online with lead times on the order of minutes. This reduces the time for predictions to deviate from reality, hence reducing the risk of inefficient decisions. 
\end{itemize}

Without fuel-based generators, reserves must be supplied by storage to ensure power balance in the face of contingencies, load variations, and renewable intermittency. However, holding slack storage capacity does not incur costs in the way that fuel-based generator reserves do.
\begin{itemize}
\item Assuming the storage's leakage is low (which is realistic for technologies like batteries and pumped hydro~\cite{Castillo2014survey}), keeping a storage device on standby is essentially free. On the other hand, fuel-based generators are expensive to keep on standby due to fuel consumption and other operating expenses.
\item Fuel-based generators incur opportunity costs because providing reserves limits the base-load power they can sell in energy markets. Storage incurs no such opportunity cost since it does not provide base-load power.
\end{itemize}

\subsubsection{Load-shifting}\label{LS}
Load-shifting refers to using storage to extract power from the system (typically at times of low net demand) and inject power later (typically at times of high net demand).
\begin{equation}\label{eq:storage0}
s_{t+1}=\alpha s_t +\Delta\left(\eta_{\textrm{in}}\max\{u_t,0\}+\eta_{\textrm{ex}}\min\{u_t,0\}\right),
\end{equation}
where $s_t$ and $u_t$ are respectively the state of charge and power injection/extraction in time period $t$, $\alpha\in[0,1]$ is the leakage over one time period, and $\eta_{\textrm{in}}\in[0,1]$ and $\eta_{\textrm{ex}}\in[1,\infty)$ are the injection and extraction losses, and $\Delta$ is the length of each time period. Typically, the state of charge, injection, and extraction are subject to energy and power capacity constraints of the form
\begin{equation} \label{eq:storage1}
0\leq s_t\leq C,\quad \left|u_t\right|\leq R,
\end{equation}
Observe that the above model only contains linear constraints. If the storage has a power-electronic grid interface and is to be used for voltage support, reactive power can be included via the convex quadratic constraint
\begin{equation}\label{eq:storage2}
u_t^2+q_t^2\leq R^2,
\end{equation}
where $q_t$ is the reactive power into or out of the storage device. Since this model only contains convex constraints, load shifting is a fairly tractable problem that can be addressed via dynamic programming~\cite{taylor2011store} or multiperiod optimal power flow ~\cite{gayme2012storage}; we discuss this further in Section~\ref{dispatch}. 

With the storage and power-electronics models outlined above, we can now dwell on the following generic optimal load-shifting problem.
\begin{equation}
\begin{array}{rl}
\underset{p,u}{\mathrm{minimize}} &\displaystyle \sum_{t=1}^T \sum_{i\in\mathcal{G}} f_{i,t}(p_{i,t}) \\
\mathrm{such\,\,that} & u_i \;\textrm{satisfies~\eqref{eq:storage0}--\eqref{eq:storage2}, }i\in\mathcal{S},\\
& \underline{p}_{i,t}\leq p_{i,t}-u_{i,t} \leq \overline{p}_{i,t},\;i\in\mathcal{S},\\
& p\in\mathcal{P}.
\end{array}
\label{eq:LS}
\end{equation}
The objective is the sum of production costs for all generators over the time period of interest. Load-shifting schedules can span hours ($T\approx10$) to a few days ($T\approx100$). A subset of nodes, $\mathcal{S}$, have storage. The second constraint represents the offset to the net supply and demand resulting from storage injections and extractions. The last constraint requires that all power quantities, $p$, be physically feasible; for example, $\mathcal{P}$ could represent the set of feasible power injections under linearized power flow and line flow limits. Note that uncertainty does not factor into our discussion of load-shifting, and is addressed later when we discuss dispatch with multiperiod optimal power flow in Section~\ref{dispatch}.

We distinguish between three functions of load-shifting.
\begin{itemize}
\item Storage can flatten the net load seen by power producers. This leads to more efficient power production because the functions $f_{i,t}(p_{i,t})$, $i\in\mathcal{G}$ in (\ref{eq:LS}) are approximately convex and increasing~\cite{WoodWoll2}. Storage profits from this practice by arbitraging price differences over time. Without fuel-based generators, system efficiency is far less dependent on output levels, and thus the price of power may not change in time; we discuss this further in Section~\ref{econsupply}. As such, we see little benefit from load-shifting for efficiency in power systems without fuel.
\item Load shifting can move power to accommodate production capacities. This makes the second constraint in (\ref{eq:LS}) easier to satisfy, thus enlarging the feasible range of operation. Much of present-day load-shifting of this type accommodates the startup and shutdown constraints of fuel-based generators. For example, flattening load makes nuclear power plants more economical by enabling them to continuously operate near their capacity limit and avoid going offline~\cite{Chen2009progress}. In power systems with high dependence on non-dispatchable resources like wind and solar, more load-shifting is needed to match supply and demand~\cite{denholm2011grid}.
\item Load-shifting can be used to reduce peak consumption levels, in turn reducing system capacity requirements. In (\ref{eq:LS}), this means attaining feasibility without changing each $\overline{p}_{i,t}$, hence reducing component installation costs in (\ref{eq:plan}). This sort of load-shifting, also known as peak-shaving, is useful with and without fuel-based generators.
\end{itemize}
Much of the storage necessitated by renewables provides services other than load-shifting such as power balancing and regulation, which are discussed in Sections~\ref{dispatch} and \ref{stability}.

Finally, a key difference between load shifting and unit commitment is that the load-shifting decisions can be implemented in near real time. This is because storage needs only a few seconds to change its power output while a fuel-based generator may take half of a day to turn on or off. Consequently, although load-shifting schedules can span more than a day, the attendant computations and decisions can be made just prior to real-time. For instance, future schedules could be continually updated in the fashion of model predictive or receding horizon control~\cite{camacho2004model} by solving~\eqref{eq:LS} every five minutes.

\begin{points*}
Unit commitment and advance procurement of reserves are not necessary without fuel-based generators. Load-shifting is critical to accommodating non-dispatchable supply and reduce capacity requirements. Although load-shifting schedules can span days, load-shifting actions can be implemented in near real-time. Consequently, there is little (if any) need for day-ahead scheduling in power systems without fuel.
\end{points*}

\subsection{Dispatch}\label{dispatch}
Power system operators continually update device settings such as transformer tap positions, the reactive power outputs of FACTS devices, and, most importantly, the real power outputs of power producers. These dispatch decisions are made by solving an optimal power flow, which minimizes instantaneous operating costs subject to the resulting power flow satisfying various network constraints. Optimal power flow can be solved at higher physical resolutions than unit commitment because the continuous nonconvexities arising from power flow constraints admit accurate linear and convex conic approximations~\cite{low2014partI,low2014partII,taylor2015COPS}. As mentioned in Section~\ref{acdc}, DC is a viable basic paradigm of power systems without fuel, which would lead to more power transfers and slightly more tractable optimal power flow computations \cite{gan2014dc}.

A generic multi-period optimal power flow routine can be written in the form
\begin{equation}\label{eq:OPF}
\begin{array}{rl}
\underset{p,q,u,v}{\mathrm{min}} &\underset{\sigma}{\mathbb{E}}\left[\displaystyle \sum_{t=1}^T\sum_{i\in\mathcal{G}} f_{i,t}(p_{i,t},\sigma)- \sum_{i\in\mathcal{L}} d_{i,t}(p_{i,t},\sigma)  \right]\\
\mathrm{s.t.} & \{p,q,u,v\} \in\mathcal{P}.
\end{array}
\end{equation}
Here, $p$ is real power, $q$ reactive power, $v$ voltage, and $u$ represents storage as in~\eqref{eq:LS}; $T$ is the total number of time periods, e.g., 48 hours would be appropriate if multi-day wind events were an issue. The function $f_{i,t}(p_{i,t},\sigma)$ is the cost of producing $p_{i,t}$ at time $t$ if $i$ is in the set of generators, $\mathcal{G}$; and $d_{i,t}(p_{i,t},\sigma)$ is the utility of consuming $p_{i,t}$ at time $t$ if $i$ is in the set of loads, $\mathcal{L}$. The expectation is over $\sigma$, a random variable that captures, e.g., uncertainty in demand or renewable output. The constraint represents all of the system physics, including power flow and storage constraints. We remark that (\ref{eq:OPF}) is a generalization of the load-shifting problem (\ref{eq:LS}), and that load-shifting schedules can be computed alongside other dispatch decisions within multi-period optimal power flow. Within the planning problem (\ref{eq:plan}), the optimization problem (\ref{eq:OPF}) would be embedded in $d(x,y,\sigma)$ and $\mathcal{P}(x)$, albeit in a simplified form for tractability.

The dominant portion of the generator cost curves, $f_{i,t}(p_{i,t},\sigma)$, $i\in\mathcal{G}$, are the fuel costs. These functions are approximately convex and increasing because generators tend to burn fuel less efficiently at higher power output levels~\cite{WoodWoll2}. In North America, the $f_i(p_i)$ are usually based on supply functions associating prices with output-power levels, which are declared by generators prior to dispatch. Currently, the load utility curves, $d_i(p_i)$, $i\in\mathcal{L}$, are often only present in unit commitment and not optimal power flow because loads have little real-time elasticity. This is changing with the advent of demand response~\cite{kirschen2003demand,kirschen2009quant}.

System operators will seek to optimally dispatch power systems without fuel, for instance, by setting storage charging and discharging schedules, choosing the set-points of power electronic devices, and adjusting the power captured from renewables, e.g., by adjusting wind turbine blade pitch and the incidence angle of PV arrays. To accommodate large-scale production by renewables,~\eqref{eq:OPF} must be modified in the following four ways:
\begin{itemize}
\item Uncertainty from forecast errors due to renewable intermittency must be included.
\item Dynamic constraints from storage, e.g.,~\eqref{eq:storage0}--\eqref{eq:storage2}, and serial correlations in renewable intermittency must be included.
\item The presence of numerous dispatchable resources like inverters and storage in low-voltage distribution systems may necessitate decentralization.
\item New objectives are needed to replace fuel costs.
\end{itemize}
We discuss these four issues in the order listed.

\subsubsection{Uncertainty}
Renewable intermittency adds significant uncertainty to dispatch decisions in the form of forecast errors. Of course, lower levels of uncertainty from loads and contingencies (e.g., line outages) have always been present in power system operations, cf.~\cite{billinton1996power}. As mentioned in Section~\ref{setup}, wind and solar power plants can be operated such that they are dispatchable up to their maximum power outputs, which are determined by random factors like wind speed and solar intensity.

The following are a few well-known, contemporary approaches to incorporating uncertainty into optimal power flow routines. Each approach would be implemented in the context of~\eqref{eq:OPF} by suitably modifying $\sigma$ and its distribution and in the feasible set $\mathcal{P}$.
\begin{itemize}
\item The chance constraint is a tool from stochastic programming~\cite{Kall1994SP,birge1997introduction} that requires the probability of an event to be above or below a certain level, for example, that the probability a collection of renewables satisfies the net load is greater than $0.99$. While generally nonconvex, some chance constraints can be cast as linear~\cite{Kall1994SP} and second-order cone constraints~\cite{Boyd1998SOCP}, the latter of which have been used as a tractable basis for chance-constrained optimal power flow~\cite{bienstock2012chance}.
\item The scenario approach is an alternative approximation to stochastic programming in which scenarios are sampled from probability distributions and used to estimate the feasibility of chance constraints. The scenario approach has also seen wide application in power systems~\cite{Vrakopoulou2013prob}, particularly in unit commitment routines~\cite{oren2011reserve}.
\item Robust optimization is a more tractable, less descriptive approach to uncertainty modeling compared to stochastic programming, which specifies uncertain sets without probability distributions~\cite{ben2009robust}. Robust optimization has also proven useful in power systems~\cite{warrington2012affine}, again particularly with unit commitment~\cite{bertsimas2013uc}.\footnote{Recently, some theoretical connections between chance constraints, the scenario approach, and robust optimization have been established in~\cite{margellos2014chance}.}
\end{itemize}

We remark that from a computational perspective, replacing fuel-based generators with renewables entails replacing intractable discrete constraints with uncertainty. While significant theoretical and empirical work is needed to quantify this tradeoff, this hypothesis appears to be reasonable for the following reasons:
\begin{itemize}
\item Renewable uncertainty admits more modeling choices than discrete unit commitment constraints, which essentially must always be modeled as integer variables.
\item Models of uncertainty are more amenable to tractable (convex) approximations. Most approaches to combinatorial optimization retain discrete modeling.
\end{itemize} 

\subsubsection{Dynamic constraints}
Flexibility from storage and demand response is necessary to accommodate renewable intermittency. Because both resources have finite energy capacity, dynamic modeling is necessary to describe the evolution of its state of charge. This includes both load-shifting as discussed in Section~\ref{schedule} and power balancing to match supply and demand on sub-minute timescales. Because the basic storage constraints,~\eqref{eq:storage0}--\eqref{eq:storage2}, are linear and convex quadratic, optimal storage decisions can be obtained alongside standard dispatch decisions via multiperiod optimal power flow~\cite{gayme2012storage}. A multiperiod optimal power flow routine like (\ref{eq:OPF}) is a sequence of single-period optimal power flow routines that are coupled period-to-period by storage dynamics like~\eqref{eq:storage0}--\eqref{eq:storage2} and other constraints such as ramp limits on power production. Multiperiod optimal power flow has approximately $NT$ variables and $MT$ constraints, where $N$ and $M$ are the number of variables and constraints in each individual period, and $T$ is the number of periods. Despite the increased number of variables and constraints, fundamentally, multiperiod optimal power flow is no more complex than ordinary optimal power flow, e.g., a multiperiod version of a second-order cone optimal power flow is still a second-order cone program~\cite{taylor2015COPS}. A useful implication of this is that stochastic and robust optimal power flow models can be extended to the multiperiod case to accommodate serial correlations in wind speeds and incident solar irradiance.

A recent alternative to multiperiod optimal power flow is risk-limiting dispatch~\cite{varaiya2011risk,Rajagopal2013risk}, which is based on dynamic programming rather than convex optimization. Risk-limiting dispatch offers less modeling versatility than multiperiod optimal power flow, but in return produces policies instead of trajectories, which better accommodate uncertainty by specifying a decision for any realization of the system state~\cite{BertsekasDPOC}.

\subsubsection{Decentralization}
We now briefly discuss decentralization. For many years, only generators and other transmission level devices could actively change their set points in response to commands. Now, components in low-voltage distribution systems like PV inverters and small energy storage devices must actively change their set points to accommodate uncertainty from distributed renewable generation; note that this largely overlaps with the paradigm of distributed generation~\cite{Ackermann2001def,pepermans2005distributed}. Computational and communication-based limitations will likely prohibit system operators from centrally operating these devices. In this case, decentralized optimal power flow algorithms will be needed to dispatch devices with limited communications. Numerous approaches already exist, including consensus-based approaches~\cite{Zhang2014volt} and the alternating direction method of multipliers~\cite{kraning2013dynamic,Dhople-DOID-2014}. While decentralization must always entail a performance loss relative to the centralized case and has long been a mathematically challenging topic, the online computational requirements of practical decentralized algorithms are generally lower than their centralized counterparts simply because they involve breaking a large problem into smaller pieces. We revisit decentralization from the perspective of real-time control on faster timescales in Section~\ref{stability}.

\subsubsection{Dispatch criteria}\label{dispatchcriteria}
Lastly in this section, we address the fact that renewables operate at zero marginal cost, i.e., once they are built, the power they produce costs very little~\cite{morales2014integrating}. We discuss the economic implications of this feature in Section~\ref{econsupply}. After removing fuel-based generators from power systems, only the load utilities remain in the objective of multiperiod optimal power flow, (\ref{eq:OPF}), which are often not present in real-time dispatch routines. Because current power system operations are almost entirely centered around minimizing fuel consumption, alternative objectives are needed for power systems without fuel; a few possibilities are listed below. Note that we do not consider resistive losses to be a valid objective due to the low marginal cost of the power itself.

\begin{itemize}
\item Maintenance is the next most significant monetary cost that explicitly depends on how the system is dispatched. Maintenance and health monitoring in power systems are well-studied topics, e.g., for pumped hydro storage~\cite{Deane2010techno}, wind turbines~\cite{walford2006wind,Hameed2009fault}, PV and storage systems~\cite{Nelson2006fuel,Diaf2008corsica,Dhople-PV-Reliability-2012}, and transmission lines~\cite{marwali1998trans,fu2007trans}. There is also a well-developed literature on maintenance theory~\cite{jardine2013maintenance}.

\item Reliability-based objectives minimize the risk of component failures and instability~\cite{billinton1996power}. We classify reliability as a non-monetary objective because it is difficult to ascribe explicit costs to potential failures. A reliability objective might include nuanced considerations such as probabilistic failures, or could be as simple as minimizing the maximum current-to-capacity ratios across the network~\cite{taylor2011DSR}. The latter can be written as
\begin{equation}
\sum_iw_i\left(\frac{x_i}{X_i}\right)^2,\label{eq:balance}
\end{equation}
where $x_i$ and $X_i$ are the usage and capacity of component $i$, e.g., the apparent power flow through and thermal capacity limit of a transmission line or power electronic converter, and $w_i$ a positive weighting factor.

\item Power quality refers to keeping the system voltage close to its nominal profile on steady-state and transient timescales~\cite{bollen2000understanding}. Power quality affects the performance and lifetime of grid-connected devices. We also classify power quality as non-monetary because the costs of voltage fluctuations are difficult to quantify.

\item Presently, most real-time dispatch routines take the loads across the system as a fixed set of parameters. Demand response enables loads to modify their aggregate consumption in real-time in response to the system's state and needs. An aggregation of active loads could submit a bid curve to the system operator associating its range of power consumptions with varying utility levels. This could be a monetary bid, cf.~\cite{albadi2007demand}.
\end{itemize}

The first three objectives overlap significantly. For example, improved reliability should improve power quality and vice versa while reducing maintenance costs. These objectives are also closely tied to transient behaviors; for example, increasing the real power output from a solar power plant will cause more disturbances in that part of the network, in turn impacting the transient stability of the system. We further discuss this coupling in Section~\ref{stability}. As a general rule, we suggest that the objective not introduce complexities beyond the constraints needed to accurately represent the physics, e.g., an objective that must be evaluated by Monte Carlo integration would compromise the tractability of the overall dispatch routine.

\begin{points*}
Dispatching power systems without fuel will differ in four ways from current practices.
\begin{itemize}
\item Forecast errors from renewable intermittency must be accounted for with uncertainty modeling.
\item Multiperiod optimal power flow must be used to model the dynamics of storage and serial correlations in renewable inputs.
\item In some cases, distributed resources will need to be operated in a decentralized fashion.
\item New objectives are required to replace fuel costs in dispatch routines.
\end{itemize}
\end{points*}

\subsection{Economics}\label{econsupply}
In this section, we discuss the economics of supplying power with only renewables. A large and growing body of literature addresses the assimilation of renewables into the current market framework, see e.g.~\cite{morales2014integrating}. The reader is referred to~\cite{Stoft2002power,kirschen2004fundamentals} for thorough expositions of power system economics. While these techniques are effective when some of the total power is produced by fuel-based generators, we believe that there is room for new perspectives when all of the power is from renewables.

Presently, most producers in North America have several revenue streams.
\begin{itemize}
\item A bilateral contract is low risk, long-term mechanism for a producer to sell a fixed quantity of power to a consumer at a price agreed upon ahead of time.
\item Producers can sell power in spot markets at nodal prices that are the dual variables of optimal power flow with a fuel cost objective. This approach was pioneered in~\cite{schweppe1988spot} and is a conceptual application of the \emph{Fundamental Theorems of Welfare Economics}~\cite{mas1995microeconomic}. An attractive quality of nodal prices is that they are outputted by optimal power flow routines and represent the sensitivity of the objective to incremental changes in power levels. Spot markets provide competitive venues for buying and selling power, which fill the essential roles of adding liquidity and driving down prices of bilateral contracts.
\item Producers are paid for providing flexibility under the names reserves and regulation. Reserve providers are often paid through call options in which they receive capacity payments for holding idle capacity on standby and energy payments if they are actually called on~\cite{arroyo2005energy,Cramton2005sense,cramton2013capacity}. Resources are also paid for providing regulation; the form of these payments has been a topic of recent debate~\cite{ferc755,taylor2013CDP}.
\item System operators or load serving entities sometimes make long-term capacity payments to power producers for being present to bid in markets and provide power over periods of years. These payments encourage construction of new generation. As discussed at the end of Section~\ref{planscale}, capacity payments could play a large role in power systems without fuel. Note, however, that their implementation would likely require a redefinition of the notion of capacity to accommodate the random maximum outputs of renewable power sources.
\end{itemize}
Bilateral contracts and spot markets are core features of today's power systems and should be present in some form if power systems without fuel are to be competitive. Flexibility is costly and essential to the reliability of power systems without fuel and hence should also be represented in markets.

Here, we focus on spot markets because they are at the foundations of current power markets but are questionable in their current form for power systems without fuel. The core challenge lies in the fact that power systems without fuel have high fixed costs and low marginal costs, much like the internet, software, and digital media~\cite{shapiro1999information,varian2004economics}.\footnote{The issues stemming from the low marginal cost of renewable energy also resemble the nuclear power discourse of the 1950s when it was said electricity would be ``too cheap to meter,'' and consumers would only need to pay for fixed costs~\cite{gamson1989nuclear}. Of course, this never came to be, and power systems without fuel will likewise require rationing mechanisms to moderate demand and control capacity expansion requirements. A basic question is how to increase fixed connection charges so that utilities can recover sales losses due to residential PV and avoid so-called ``death spirals''~\cite{Cai2013PV}.} We discuss the economics of regulation further in Section~\ref{transientecon}.

Because there are no fuel costs, employing nodal pricing as is would simply result in all prices being equal to zero. The fuel cost objective in (\ref{eq:OPF}) could hypothetically be replaced with one or a combination of the objectives discussed in Section~\ref{dispatch} such as maintenance costs, and nodal prices could be defined as usual. In this case, any new objective in (\ref{eq:OPF}) must be monetary for the dual multipliers to have valid price interpretations; if the objective is not monetary, then the dual multipliers are simply the sensitivities of the optimal objective and not marginal cost-based prices. We must, however, ask whether this is appropriate for power systems without fuel. The following are three fundamental issues.
\begin{itemize}
\item Time-varying nodal prices reflect the instantaneous physical efficiency and hence marginal cost of operating the power system, which currently varies on a minute-to-minute basis. This results in generators being paid according to their efficiency, and, under dynamic pricing~\cite{borenstein2002dr}, incentivizes efficient power consumption over time. Although demand and system capacity will vary in time, it is not clear that the marginal cost of operating power systems without fuel varies on a minute-to-minute or even hourly basis.
\item It is straightforward to map generation levels to the cost of fuel consumption, making it relatively easy to audit bidding by fuel-based generators. The costs of maintenance, reliability, and power quality are harder to quantify and hence harder to audit. Bidding mechanisms based on these costs may consequently be more susceptible to market power and gaming.
\item Demand-side bidding alone cannot account for the cost of supplying power, but rather only what consumers are willing to pay, hence removing producer input from the price-setting process. Such an arrangement can lead to excessively small payments to producers. For example, if we model such demand-side bidding with supply function competition~\cite{klemperer1989supply,green1992british}, it is trivially a pure strategy Nash equilibrium for all bids to be zero, thus rendering zero payments.
\end{itemize}
Due to the above issues, it is our stance that replacing fuel-based generators with renewables warrants a rethinking of power markets.

\begin{points*}
The low marginal and high fixed costs of renewable power production makes the current nodal pricing framework dubious for power systems without fuel. The development of new market mechanisms will rely on accurately quantifying the cost of operating power systems without fuel on all timescales. Since these costs may be based on a number of complex and hard to audit factors, market power and gaming are important considerations.
\end{points*}

\section{Transient behavior}\label{stability}
Several challenges will need to be addressed from modeling, analysis, and control perspectives as we embrace the transition from electromechanical energy-conversion interfaces (synchronous machines) delivering power from thermal sources (coal, nuclear, gas) to semiconductor-based energy conversion interfaces (power-electronic converters) delivering power from renewable resources (wind and solar). The main challenges in this context are the following.
\begin{itemize}
\item The system dynamics in the envisioned power systems without fuel are not dominated by mechanical generator rotor inertias. Apart from electrochemical and electromechanical energy storage devices (batteries and flywheels, respectively) power-electronic energy conversion interfaces offer limited inertia innately.
\item Steady-state dispatch decisions now have greater effect on transient stability. For instance, procuring a large amount of power from a PV power plant also increases the magnitude of the disturbances it contributes.
\item Dynamics of fast-switching power converters present imposing analysis challenges. Quasi-stationary phasor representations of electrical-network voltages and currents that are commonly used in bulk power systems lack the required modeling fidelity in power systems without fuel.
\end{itemize}

While one could argue that with the reduced mechanical inertia, a no-fuel power system is likely to be more sensitive to disturbances, conceivably, there is also more control flexibility since power-electronic interfaces can act on much faster timescales compared to synchronous generators. Next, we offer perspectives on how modeling, analysis, and control frameworks will need to be revised in a no-fuel future.

\subsection{Modeling}
Problems related to stability in power systems dominated by synchronous generators have received significant attention over the years. A classification of different stability notions and links to reliability and security is available in the report~\cite{Kundur-2004}. From the generation perspective, stability issues are typically studied for a networked collection of synchronous generators, with nodal dynamics modeled by the ubiquitous \emph{swing equations}. Adopting a standard one-axis model for synchronous generators in the power system~\cite{Sauer-1990}, the swing equations are given by 
\begin{align} \label{eq:syngenswing}
\frac{\mathrm{d}\delta}{\mathrm{d}t}&=\omega-\omega_\mathrm{synch}, \nonumber\\
\frac{\mathrm{d}\omega}{\mathrm{d}t}&=\frac{1}{M}\left(P_\mathrm{mech}-P_\mathrm{elec}-D(\omega-\omega_\mathrm{synch})\right),
\end{align}
where $\delta$ and $\omega$ are the rotor angle and speed, $\omega_\mathrm{synch}$ is the synchronous speed, $M$ and $D$ are the inertia constant and damping coefficients, $P_\mathrm{mech}$ is the mechanical-input power, and $P_\mathrm{elec}$ is the electrical power that flows into the network.

Dynamic models for power-electronic inverters are (circuit) topology and application dependent, and difficult to abstract in a convenient analytical form akin to the ubiquitous swing equations. This is particularly due to the following reasons:
\begin{itemize}
\item digital control methods that are typically employed in power-electronics circuits involve state-machine-type or look-up-table-based algorithms that are difficult to encode as dynamical systems;
\item the utilization of switching power semiconductor devices in power electronic systems implies that switching-time-scale behavior is only well modeled by hybrid automata for which scalable simulation platforms are lacking; 
\item characteristics of renewable power sources and energy storage devices are nonlinear, technology specific, and use proprietary control algorithms that are not readily accessible in the literature.
\end{itemize}
The challenges mentioned above have been tackled with modeling strategies that include: virtual prototyping~\cite{nguyen2014tec}, reachability analysis~\cite{Hope-2011}, and sampled data modeling and control for power-electronics circuits~\cite{Kikuchi-2012,Almer-2013,Goodwin-2013}. 

\subsection{Analysis}
In their original form, dynamical models that capture the electromechanical behaviors of synchronous generators in a power network are described by differential algebraic equations. Algebraic equations are introduced from real- and reactive-power-balance conditions for nodes in the electrical network. Given the lack of numerical-simulation, stability-analysis, and controller-design methods for DAE models, it is common to resort to reduced models of the electrical network when analyzing a collection of synchronous generators. These reduced models are typically composed of nonlinear ODEs and are recovered by eliminating the algebraic power-flow equations. A variety of electrical-network model reduction methods have been developed for bulk power systems~\cite{Chow-2013}. To illustrate the limitations of existing approaches and outline future directions, we focus the subsequent discussion (without loss of generality) on \emph{Kron reduction}. 

Given an electrical network with a set of pre-specified terminals, Kron reduction recovers an electrically equivalent circuit retaining terminal nodes and eliminating non-terminal nodes~\cite{Kron-1939,Dorfler-13}. Kron reduction is accomplished by taking the Schur complement of the admittance matrix. In general, encoding network interactions through the admittance matrix is only possible when we rely on quasi-stationary sinusoidal-steady-state representations for the electrical network variables~\cite{schiffer2013synchronization,schiffer2013stability,BG-JWSP-FD-SZ-FB:13a,Porco-2012-1,JWSP-FD-FB:12u}. However, such quasistationary models admittedly lack the fidelity required to capture network transient behaviors at faster than AC-cycle time scales. Therefore, with regard to analysis of large networks of interconnected inverters, there is a pressing need to develop scalable model-reduction methods that can be applied across a wide spectrum of time scales. Noteworthy efforts in this direction are time-domain Kron reduction methods~\cite{SYC-PT:12,Caliskan-2012}.

\subsection{Feedback Control}
Frequency across the bulk power system is, to the first order, backed up by the mechanical inertia on offer from the synchronous generators. In particular, notice from~\eqref{eq:syngenswing} that variations of frequency to load fluctuations, i.e., $P_\mathrm{elec}$ are inversely related to the generator mechanical rotor inertia. As such, power imbalances in synchronous generators directly translate to variations in electrical frequency. This fundamental relationship underlies turbine-governor control (at the generator level) and load frequency control (at the system level). In a future power system dominated by power-electronics interfaces, mechanical inertias will no longer be direct proxies for nodal frequency dynamics across the electrical network. The limited inertia on offer in power-electronics inverters is that provided by inductive and capacitive filter elements; and therefore  without fuel-based generators, electrical physics can be expected to dominate at even faster timescales. The role of feedback control in this setting is magnified, and next-generation distributed-control algorithms will play an important role of maintaining stable voltages and frequencies in a network of interconnected power electronic inverters. 

In the present generation of power-electronics inverters intended for grid-tied operation, controllers are designed with the implicit assumption of the existence of a stiff voltage source. The main role of inverters in this setting is to regulate the delivery of real and reactive power. This is the so-called \emph{grid-following} control paradigm. Recognizing the inevitable transition to a network with a high penetration of power-electronics inverters, there has been resurgent interest in control methods that seek to regulate terminal voltages and frequencies in the absence of a stiff grid voltage. These approaches have been referred to as \emph{grid-forming} controls. Efforts in this direction are largely based on the so-called droop control~\cite{Chandorkar-1993}. Inspired by the operation of synchronous generators in bulk power systems, the premise of droop control is to linearly trade off terminal voltage and frequency versus active and reactive power. To leverage the increased control capabilities offered in fast-acting power-electronics inverters, recently there has been some interest in the development of time-domain control methods that directly act on time-domain waveforms and do not assume sinusoidal steady-state operation as is done in droop control. System-theoretic methods have outlined conditions under which frequency-synchronized electrical variables can be maintained across the network with droop control~\cite{Chandorkar-1993} and time-domain approaches~\cite{TorresHespanhaMoehlisJul12,SD-SynchTPELS-2014}. Extending these analytical methods to accommodate a variety of energy-conversion interfaces as well as fuel-based synchronous generators is, albeit challenging, critical to understand the implications of high penetrations of power-electronics inverters.

\subsection{Coupling between the transient and steady-state timescales}
The spatiotemporal characteristics of power systems without fuel are fundamentally different from present bulk power systems. In particular, the generation and load are closely collocated, largely intermittent, and vary at faster timescales. This aspect motivates the development of new approaches to system management: hierarchical operational practices that enforce a strict time-scale separation between energy management and real-time control are incompatible with the form and function of no-fuel power systems. 

Traditionally, network-wide energy management, e.g., optimal power flow and load-shifting, has been designed independently from feedback control mechanisms which track set-points from energy management systems, e.g., automatic generation control and droop control. This approach is based on the separation between steady-state (seconds to minutes) and transient (sub-cycle to seconds) timescales. In power systems without fuel, adopting this time-scale separation to guide optimization and control is questionable for the following reasons:
\begin{itemize}
\item Timescale separation is merely a simplifying approximation; designing energy management and feedback mechanisms with regard to both timescales will result in better performance~\cite{FD-JWSP-FB:14a}. Moreover, increasing the frequency at which energy management is executed will also improve performance. Improved computational capabilities make such unified designs viable.
\item The dispatch of variable wind and solar power sources strongly influences the location and magnitude of disturbances, in turn influencing the system's transient behaviors and regulation needs. Hence, decisions at the steady-state timescale strongly influence performance at the transient timescale.
\end{itemize}
In this section, we discuss the implications of jointly addressing the steady-state and transient timescales.

\subsubsection{Feedback control and energy management}
The goal of feedback control is to maintain stable AC waveforms (uniform voltages and frequency) across the network in spite of variations in sources, loads, and exogenous disturbances. The objective of energy management is to provide set-points to the real-time inverter controllers such that in steady-state, power losses and voltage deviations are minimized and economic benefits to end users are maximized. 

Several approaches can therefore be adopted to steer the operating point of a power system without fuel to an energy management routine's solution. One solution strategy that has received significant attention in the literature is to synthesize controllers that seek saddle points of Lagrangian functions corresponding to the network-optimization problems, and then integrating these optimization-based controllers with existing real-time controllers. The general strategy of synthesizing dynamical systems to solve convex optimization problems goes back to the seminal work of~\cite{Arrow-1958}, and in recent years, it has received increased attention in the control systems community in light of several motivating applications~\cite{HirataHespanhaUchidaJun14,Durr-2012,Brunner-2012,Jokic-2008,Jokic-2009,DeHaan-2005,NaLi-2014,Chen-2014}. While many synthesis approaches have been put forth, one that is conceptually straightforward is to formulate dynamical systems (the states of which are proxies for Lagrange multipliers) which evolve in a gradient-ascent-like fashion towards the optimal dual solution~\cite{Jokic-2009}. Reference inputs for the physical-layer system are then derived from these optimization-based dynamical states. Since these controllers are formulated to dynamically seek the Karush-Kuhn-Tucker (KKT) conditions for optimality, they have been referred to as \emph{dynamic KKT controllers}~\cite{Jokic-2008,Jokic-2009}.


\subsubsection{Economics}\label{transientecon}
Steady-state dispatch influences the location of disturbances in power systems without fuel, which in turn influences which resources will be tapped for regulation. The fast adjustments involved in providing regulation, e.g., rapidly cycling the state of charge of a battery, factor significantly into maintenance costs. As discussed in Sections~\ref{dispatchcriteria} and \ref{econsupply}, these maintenance costs may play a significant role in the steady-state dispatch and economics of power systems without fuel. Hence, not only must energy management and feedback control mechanisms be jointly designed, but economic mechanisms must also take into account costs incurred on transient time scales.

We now comment on who should bear the cost of regulation. Historically, most disturbances have been the result of natural load variations and component failures. Load-following handles the first type of disturbance and should be paid for by the loads. Failures, however, are the result of machine breakdowns and exogenous factors like trees and lightning. Failures are only problematic on fast timescales because loads require uninterrupted power. Consequently, it is the responsibility of both the component owner and loads to pay the cost of failures, cf.~\cite{Strbac200032,kirby2000customer}. Such costs are often termed negative externalities because they are not clearly attributable to the actions of a specific market participant. In his seminal work, Coase identified such scenarios as problems of social cost and concluded that an appropriate division of costs can be obtained by negotiation between all affected market participants~\cite{coase1960problem}. Disturbances from variable renewable power production are also negative externalities because the incurred cost would be zero either if renewables did not create disturbances or if loads did not require steady power. Hence, the cost of disturbances from renewables should be borne by both renewables and the loads they serve. The design of such cost allocation mechanisms is an important research topic in power system economics.

\begin{points*}
Adoption of semiconductor-based power-electronics interfaces in power systems without fuel will necessitate development of new modeling, analysis, and control schemes. The following are some specific challenges.
\begin{itemize}
\item Inverter controllers will evolve to realize \emph{grid-forming} controllable voltage sources as compared to \emph{grid-following} current sources.
\item Management schemes that bridge the temporal gap between real-time control and energy management are necessary to realize the full potential of responsive power-electronics interfaces.
\item Maintenance of resources providing fast regulation may represent a major operational cost in power systems without fuel. New economic mechanisms may be necessary to represent these costs in markets on slower timescales.
\end{itemize}
\end{points*}

\section{Summary}\label{conclusion}
Climate change and the finiteness of fossil fuels make it viable that many future power systems could run entirely without fuel. Considerable technical and economic challenges lie between present-day power systems and power systems without fuel. Foremost, the intermittency of renewables entails that new sources of flexibility like storage and demand response must become more cost-effective for removing all fuel-based generation to be feasible. A new array of optimization and control algorithms must then be developed to ensure the safe and efficient operation of power systems without fuel. Complementary economic mechanisms that incentivize proper behavior and investment and which are robust to market abuse must also be developed. While daunting, we regard these challenges to be surmountable through research, technological development, and policy.

We believe that power systems without fuel will differ substantially from fuel-based power systems beyond environmental benevolence and sustainability. Many of these differences stem from the facts that all timescales are shrinking, and that there will be numerous small components instead of a few large components. Some of these differences translate to genuine advantages, including closer to real-time decision-making with more accurate information and the viability of a predominantly DC infrastructure, to name a few.

\section*{Acknowledgments}
J. A. Taylor would like to thank Pravin Varaiya, Kameshwar Poolla, Peter Lehn, and Reza Iravani for helpful discussion.
\bibliographystyle{IEEEtran}
\bibliography{MainBib,Edited_References,Ref1}

\end{document}